# RVS SPECTRA OF GAIA PHOTOMETRIC SCIENCE ALERTS

**George SEABROKE[1], Mark CROPPER[1], Steven BAKER[1], Kevin BENSON[1], Chris DOLDING[1], Mike SMITH[1], Arancha DELGADO[2], Simon HODGKIN[2], Diana HARRISON[2], Guy RIXON[2], Lukasz WYRZYKOWSKİ[3]**

[1] *University College London, Faculty of Mathematical & Physical Sciences, Department of Space & Climate Physics, Mullard Space Science Laboratory, Holmbury St. Mary, Dorking, Surrey, RH5 6NT, United Kingdom (email: g.seabroke@ucl.ac.uk).*

[2] *University of Cambridge, Faculty of Physics and Chemistry, Institute of Astronomy, Madingley Road, Cambridge, CB3 0HA, United Kingdom.*

[3] *Astronomical Observatory, University of Warsaw, Al. Ujazdowskie 4, 00-478 Warszawa, Poland.*

**Abstract:** Gaia Photometric Science Alerts (GPSA) publishes Gaia *G* magnitudes and Blue Photometer (BP) and Red Photometer (RP) low-resolution epoch spectra of transient events. 27 high-resolution spectra from Gaia's Radial Velocity Spectrometer (RVS) of 12 GPSAs have also been published. These 27 RVS epoch spectra are presented next to their corresponding BP and RP epoch spectra in a single place for the first time. We also present one new RVS spectrum of a 13th GPSA that could not be published by the GPSA system. Of the 13 GPSA with RVS spectra, five are photometrically classified as unknown, five as supernovae (three as SN Ia, one as SN II, one as SN IIP), one as a cataclysmic variable, one as a binary microlensing event and one as a young stellar object. The five GPSAs classified as unknown are potential scientific opportunities, while all of them are a preview of the epoch RVS spectra that will be published in Gaia's fourth data release.

## 1. Introduction

It is well known that Gaia's Radial Velocity Spectrometer (RVS, Cropper et al. 2018) contributed the largest ever radial velocity catalogue of more than 7 million stars (Sartoretti et al. 2018, Katz et al. 2018) to Gaia's second data release (DR2, Brown et al. 2018). It should also be known that RVS spectra of spectroscopically well-behaved objects, for which object classification and astrophysical parameters can be derived, will be published in Gaia's third data release (expected in the second half of 2021). What is less well known is that 27 RVS spectra of 13 Gaia Photometric Science Alerts (GPSAs, Campbell et al. 2014) have already been published (see Table 1).

"[GPSA] is an all-sky photometric transient survey, based on the repeated, high-precision measurements made by the Gaia satellite. … The Cambridge Institute of Astronomy DPCI runs a dedicated data processing pipeline to look for transient events in the Gaia data, i.e. "new" sources where previously nothing was detected, or sudden, dramatic changes in brightness of previously detected stars. [GPSA] are made public immediately after the data processing and alert identification, typically 2-3 days after the observation by the satellite. The dedicated per-alert pages include all the data Gaia has collected for that source, including the light curves and low-resolution BP/RP spectra."[1]

The aim of this paper is to advertise the existence of RVS spectra of GPSAs, linked to the per-alert pages, present a new RVS spectrum that that could not be published by the GPSA system, highlight the potential scientific opportunities of the five unclassified GPSAs with RVS spectra and present all the RVS spectra of GPSAs as a preview of the epoch RVS spectra that will be published in Gaia's fourth data release.

[1] http://gsaweb.ast.cam.ac.uk/alerts/about

[2] http://gsaweb.ast.cam.ac.uk/alerts/alertsindex

**Table 1**: GPSA index page[2] after clicking on the RVS column header to display those with RVS spectra. The line for Gaia17aro is also displayed.

| Name | TNS | Observed | RA (deg.) | Dec. (deg.) | Mag. | Historic mag. | Historic scatter | Class | Published | Comment | RVS |
|---|---|---|---|---|---|---|---|---|---|---|---|
| Gaia17cke | AT2017gyv | 2017-09-22 13:28:57 | 262.63035 | -32.70416 | 16.76 | 18.53 | 0.32 | CV | 2017-09-30 08:35:55 | red Gaia source towards Galactic centre brightens by >2 mags; strong emission lines in BP/RP | ✔ |
| Gaia17bxg | SN2017erp | 2017-07-27 14:19:42 | 227.31168 | -11.33418 | 14.59 | | | SN Ia | 2017-07-31 08:26:15 | bright confirmed SN Ia, SN 2017erp, in Seyfert 2 galaxy NGC 5861 | ✔ |
| Gaia17axt | SN2017bzc | 2017-04-04 01:10:21 | 349.06129 | -42.56960 | 13.33 | | | SN Ia | 2017-04-05 13:00:29 | very bright candidate SN in outskirts of galaxy APMBGC 291+064-120, aka AT 2017bzc | ✔ |
| Gaia16bgk | SN2016fej | 2016-09-11 16:21:53 | 310.16636 | -54.31064 | 14.15 | | | SN Ia | 2016-09-14 10:57:31 | candidate SN in NGC 6942 GS-TEC predicts SN Ia | ✔ |
| Gaia17axu | SN2017bzb | 2017-04-03 07:06:34 | 344.32216 | -41.01596 | 13.08 | | | SN II | 2017-04-05 13:03:37 | very bright candidate SN in outskirts of galaxy LCRS B225429.0-411654, aka AT 2017bzb | ✔ |
| Gaia17bmy | SN2017eaw | 2017-06-08 22:38:47 | 308.68432 | 60.19329 | 12.87 | | | SN IIP | 2017-06-12 11:58:24 | very bright confirmed SN IIP, in galaxy NGC 6946, aka SN2017eaw | ✔ |
| Gaia16aye | AT2016eta | 2016-08-05 00:53:52 | 295.00474 | 30.13149 | 14.27 | 15.51 | 0.06 | ULENS | 2016-08-09 10:45:38 | 1.2 mag rise in red star near Galactic Plane | ✔ |
| Gaia17azz | AT2017dbx | 2017-04-13 21:02:36 | 105.88408 | -6.62354 | 16.75 | 18.08 | 0.59 | unknown | 2017-04-16 08:15:04 | >1 mag brightening of Young Stellar Object candidate | ✔ |
| Gaia17cmj | AT2017haw | 2017-09-24 13:26:26 | 273.96540 | -31.11752 | 13.12 | 12.07 | 0.20 | unknown | 2017-10-03 11:03:12 | W Vir star fades by 1 mag and becomes redder | ✔ |
| Gaia16bnz | AT2016hjn | 2016-10-17 12:59:20 | 55.07493 | 49.35893 | 13.71 | 12.70 | 0.19 | unknown | 2016-10-24 13:35:18 | 1.3 mag decline in 13th mag blue source over >1yr | ✔ |
| Gaia16bpa | AT2016hmx | 2016-10-18 04:31:25 | 272.45945 | -21.76201 | 16.47 | 15.40 | 0.13 | unknown | 2016-10-27 08:47:25 | 1 mag decline in red star in Galactic plane | ✔ |
| Gaia17aeq | AT2017md | 2017-01-19 02:42:10 | 77.54588 | -3.47394 | 15.68 | 13.49 | 0.18 | YSO | 2017-01-20 21:05:02 | long term decline in brightness, near YSO 2MASS J05101100-0328262 (0.11 arcsec), and ASASSN -13db | ✔ |
| Gaia17aro | AT2017cfx | 2017-03-12 21:07:43 | 112.65553 | -22.31571 | 17.16 | 18.07 | 0.19 | unknown | 2017-03-18 14:47:35 | long-term increase in brightness of red Gaia source in galactic plane, past variability, cand. YSO | |

## 2. Method

The UK Gaia grant funds the Cambridge Institute of Astronomy to develop the GPSA pipeline and to run it. Between 2015-2018, it also funded the Mullard Space Science Laboratory to develop a dedicated data processing pipeline to provide RVS spectra of GPSA. The pipeline is a bespoke version of the one used to produce the radial velocities in Gaia DR2 (Sartoretti et al. 2018).

Each per-alert page has a table of photometric epochs. Clicking on a row displays the corresponding Blue Photometer (BP) and Red Photometer (RP) spectra. If there is one or more corresponding RVS spectrum, there is an *RVS spectra* column in the table. The epoch with a RVS spectrum has a link called *spectrum*, which opens a pop-up window displaying the RVS spectrum. Within this window, there is a link called *RVS data reduction Info*, where the processing is described.

## 3. Results

GPSAs are provided with a *Class* column (see Table 1), giving the type of transient event. The five GPSAs classified as unknown are plotted in Figures 1-5, the binary microlensing event in Figure 6, the CV in Figure 7, the three SN Ia in Figure 8, the SN II in Figure 9, the SN IIP in Figure 10 and the YSO in Figure 11. Each figure includes a light curve delineating which epochs have RVS spectra and the epoch RVS spectra next to their corresponding BP/RP spectra. Figures 1 and 2 display flat-line BP/RP spectra, which denotes a problem with those spectra during that epoch.

[2] http://gsaweb.ast.cam.ac.uk/alerts/alertsindex

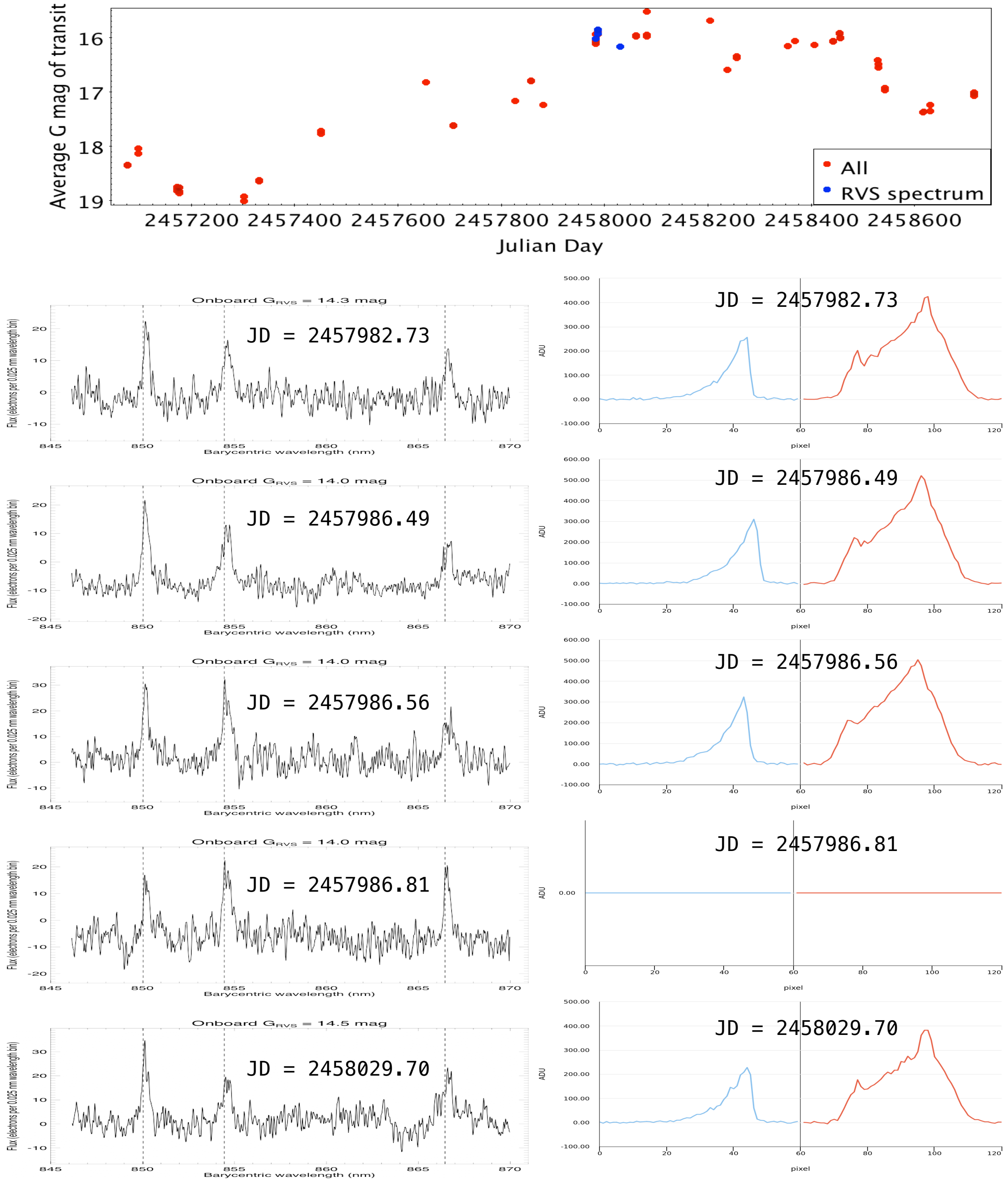

**Figure 1**: Gaia17azz (*Class*: unknown) light curve (top), RVS spectra (left, black lines) and BP spectra (middle, blue lines) and RP spectra (right, red lines). Vertical dotted lines in the RVS spectra denote the wavelengths of the Ca triplet lines.

## 4. Discussion

Gaia17azz (*Class*: unknown): The comment on the alert page is ">1 mag brightening of Young Stellar Object [YSO] candidate". The YSO conjecture is supported by all the RVS spectra exhibiting their calcium lines in emission (Figure 1). The calcium emission lines may be exhibiting evolution in the both their strengths and shapes.

For example, the bluest calcium emission line peaks at about 20 electrons per 0.025 nm wavelength bin in the first two epochs (separated by about four days), which increases to 30 in the third epoch two hours later, back to 20 in the fourth epoch six hours later, then back to 30 in the fifth epoch 43 days later.

Gaia16bpa (*Class*: unknown): The comment on the alert page is "1 mag decline in red star in Galactic plane". The light curve in Figure 2 shows this decline was very short lived. The RVS spectrum observed immediately after the short-lived decline in brightness has weak emission in its calcium lines. This emission is not apparent in the following two epochs, suggesting the emission may be linked with the decline in brightness.

Gaia17cmj (*Class*: unknown): The comment on the alert page is "W Vir star fades by 1 mag and becomes redder". The two RVS spectra in Figure 3 were observed after the fading and do not appear to exhibit any unusual features.

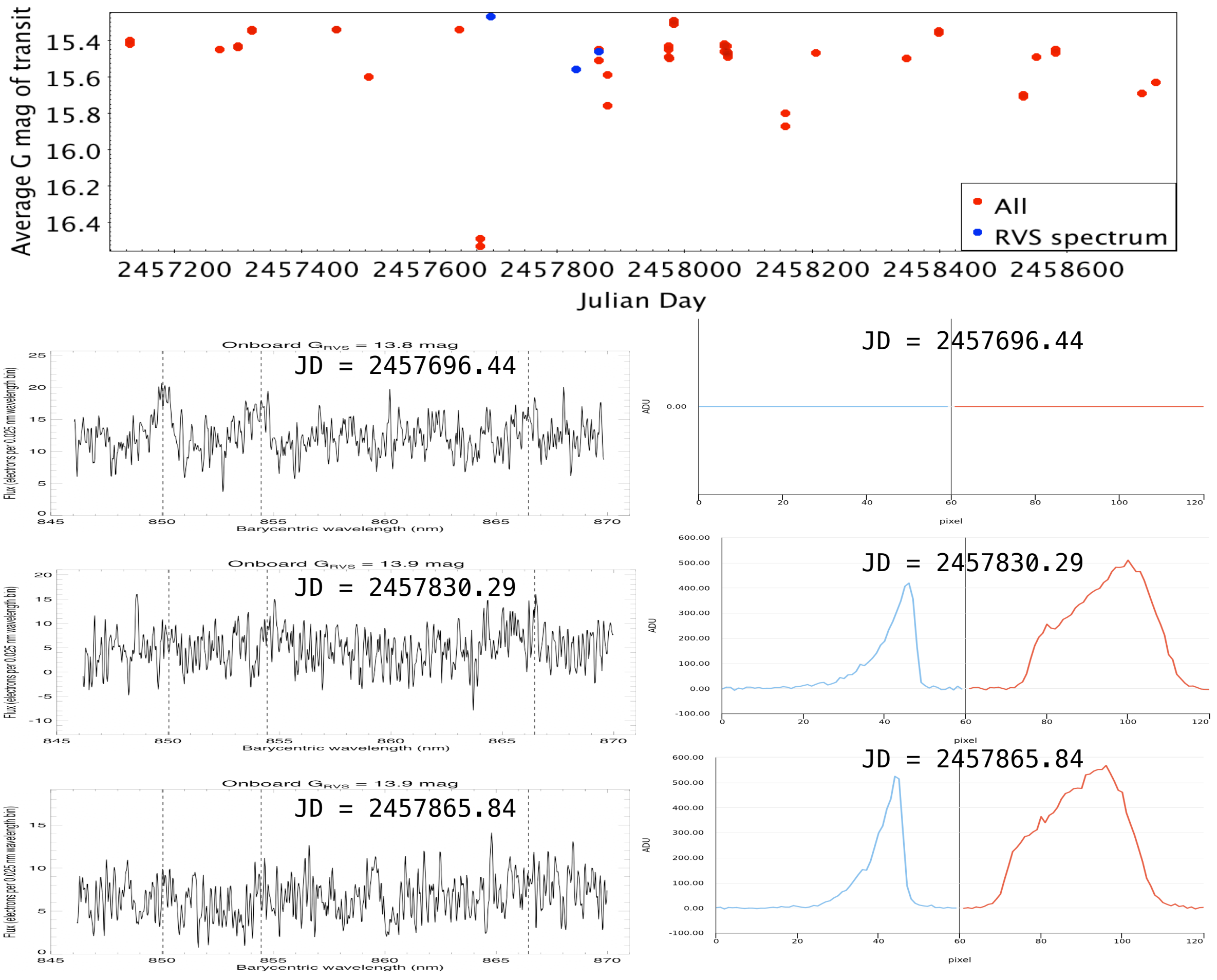

**Figure 2**: Same as Figure 1 but for Gaia16bpa (*Class*: unknown).

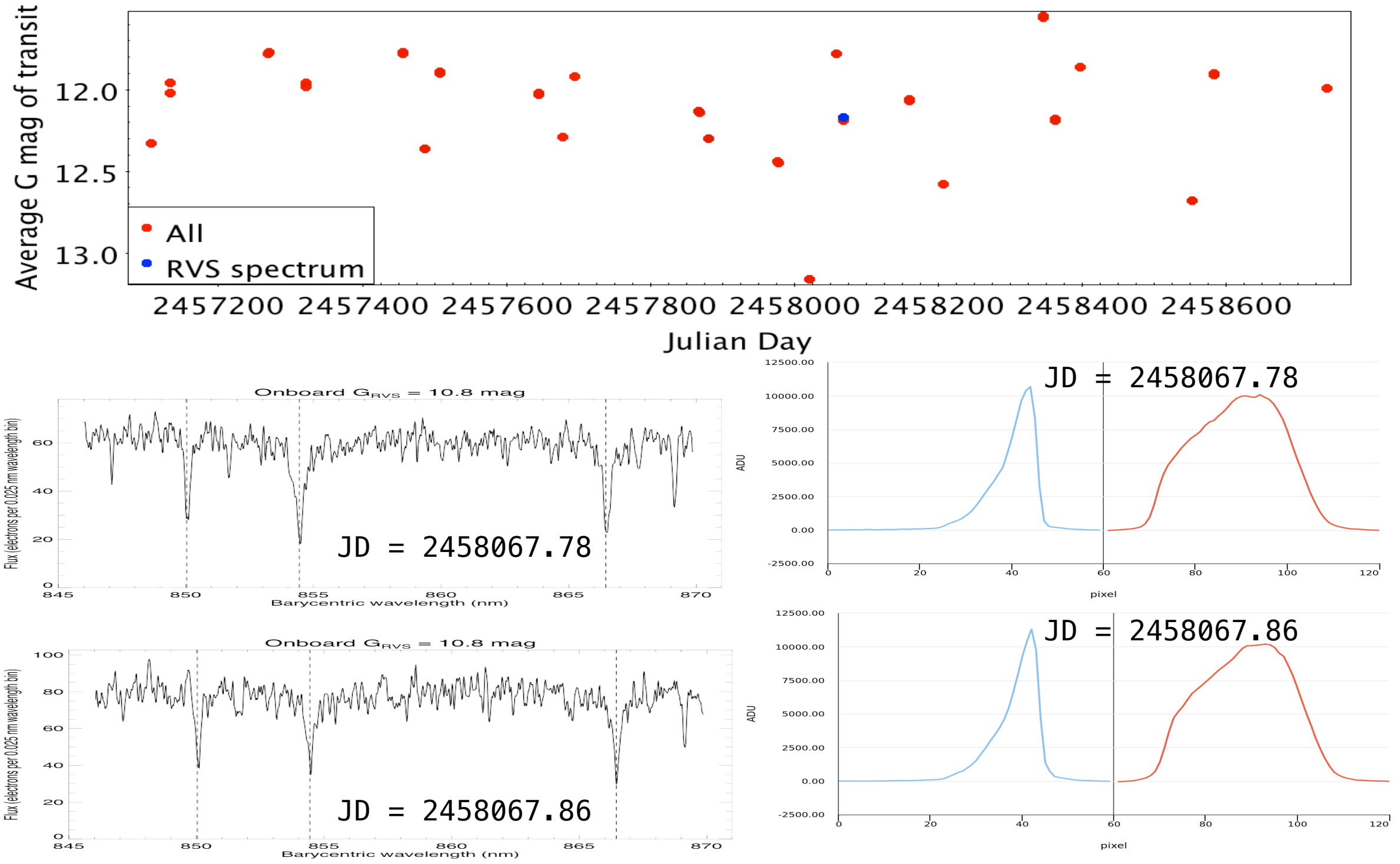

**Figure 3**: Same as Figure 1 but for Gaia17cmj (*Class*: unknown).

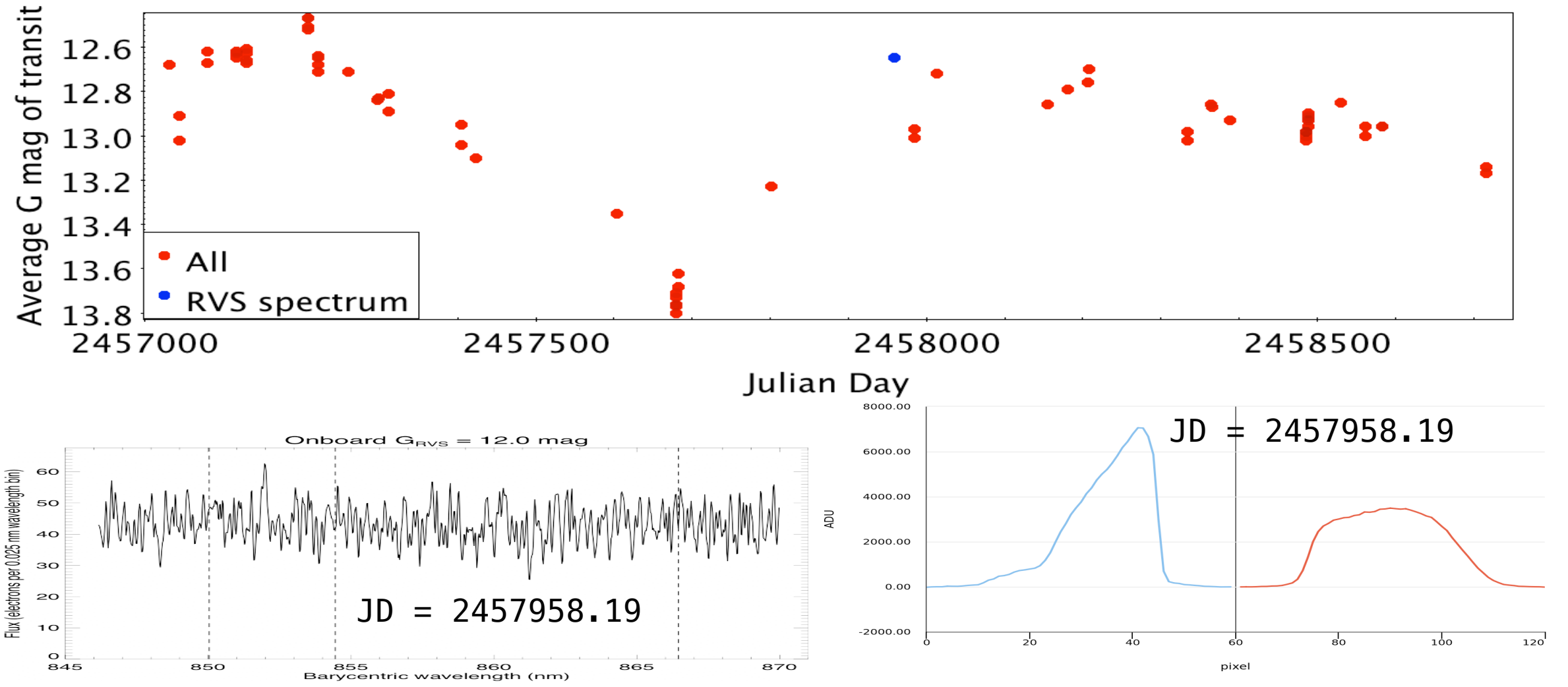

**Figure 4**: Same as Figure 1 but for Gaia16bnz (*Class*: unknown).

<u>Gaia16bnz (*Class*: unknown):</u> The comment on the alert page is “1.3 mag decline in 13th mag blue source over >1yr”. It has been suggested that the light curve for this source is nova-like, possibly of VY Sculptoris type[3]. The light curve in Figure 4 shows that the RVS spectrum was observed during no obscuration. The featureless continuum is consistent with a white dwarf type DC if this is very short period system.

[3] http://ooruri.kusastro.kyoto-u.ac.jp/mailarchive/vsnet-campaign-nl/351

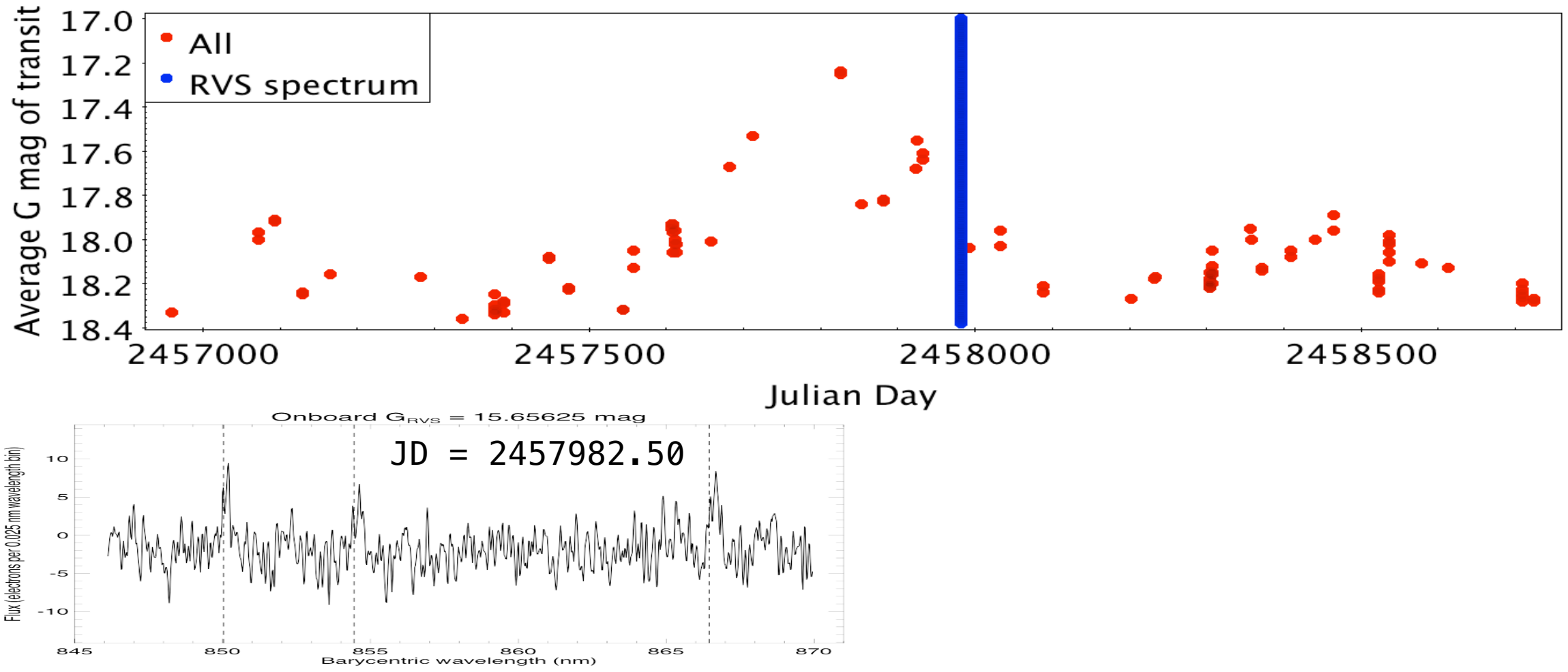

**Figure 5**: Gaia17aro (*Class*: unknown) light curve (top) and RVS spectrum (left, black line). The epoch of the RVS spectrum does not correspond to any photometry in the light curve so the RVS spectrum epoch is denoted as a vertical line in the light curve. Also this epoch also does not include a BP or RP spectrum.

Gaia17aro (*Class*: unknown): The comment on the alert page is "long-term increase in brightness of red Gaia source in galactic plane, past variability, candidate YSO". The calcium emission lines in the RVS spectrum in Figure 5 support this. Its RVS spectrum could not be published by the GPSA system so Figure 5 presents this spectrum for the first time.

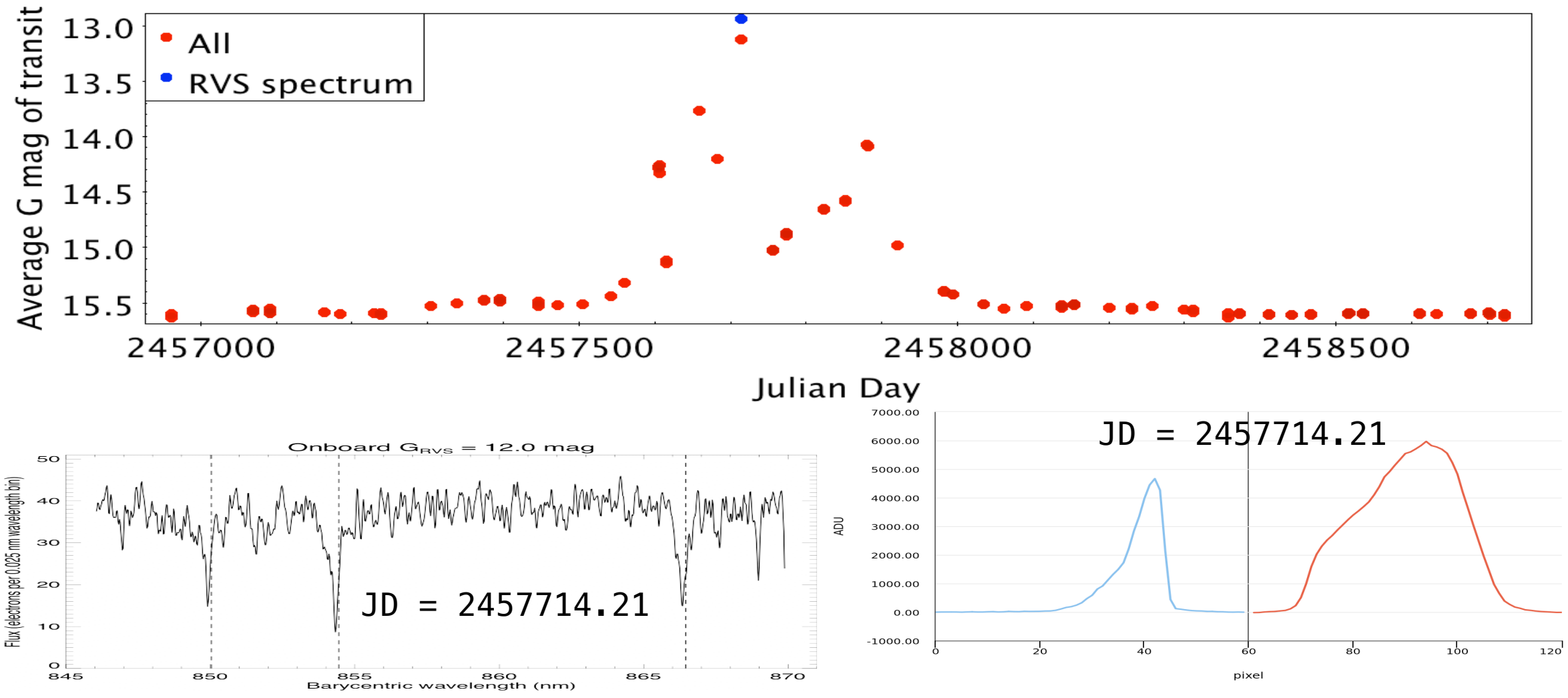

**Figure 6**: Same as Figure 1 but for Gaia16aye (*Class*: ULENS).

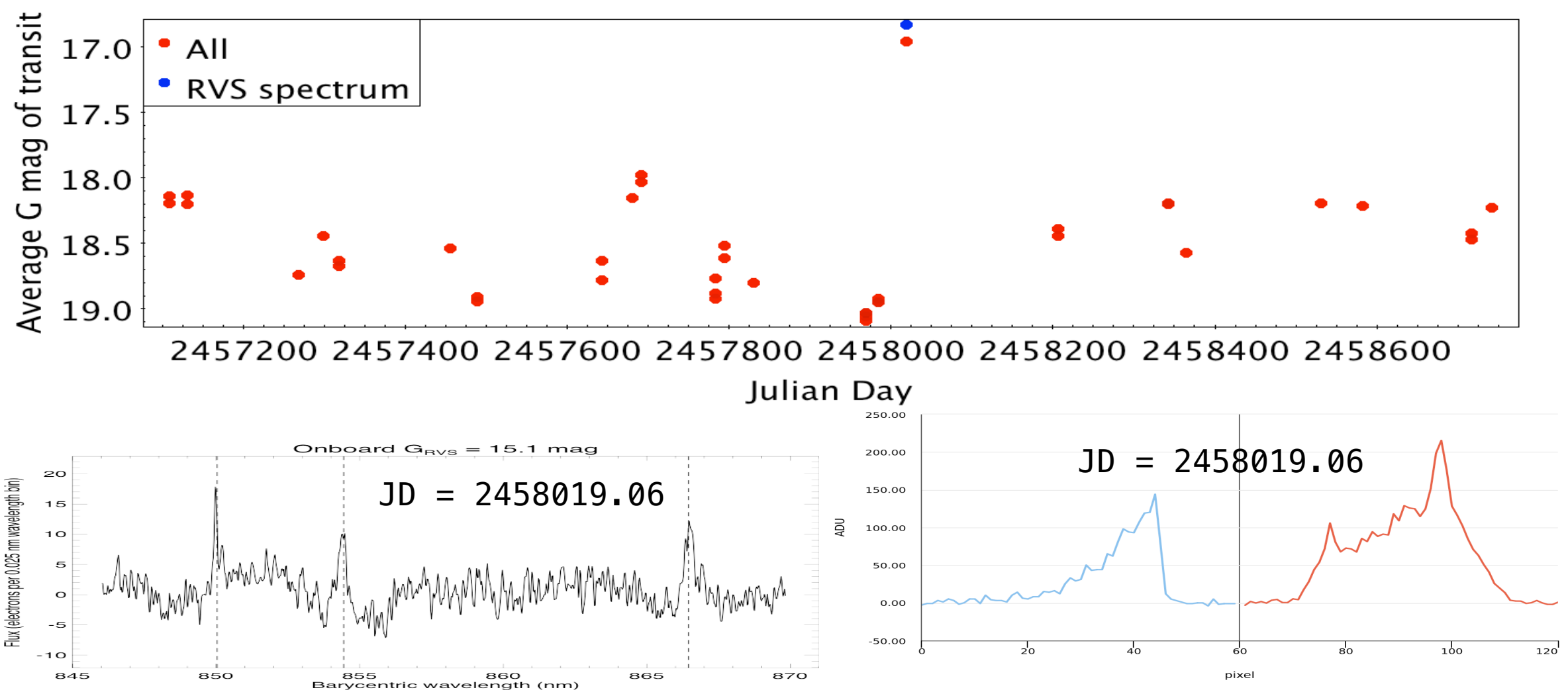

**Figure 7**: Same as Figure 1 but for Gaia17cke (*Class*: CV).

Gaia16aye (*Class*: ULENS): The RVS spectrum in Figure 6 is the only one from GPSAs to be included in a published paper (Wyrzykowski et al. 2019). The RVS spectrum is from the lensed source at the brightest moment of the event as seen by Gaia at the fourth caustic crossing.

Gaia17cke (*Class*: CV): The comment on the alert page is "red Gaia source towards Galactic centre brightens by >2 mags; strong emission lines in BP/RP". The RVS spectrum in Figure 7 also exhibits calcium emission, which is also seen in CVs (Cropper & Marsh 2003). The broad absorption lines at 848 and 855 nm may be part of the Paschen series but the Paschen lines at 860 and 867 nm, which are expected to be stronger than the 848 and 855 nm ones, are not visible.

Gaia17bxg, Gaia17axt, Gaia16bgk, Gaia17axu, Gaia17bmy (*Class*: SN): The RVS spectra of supernovae in Figures 8, 9 and 10 exhibit continuum shapes and some broad features, presumably the calcium triplet lines appearing as one broad absorption line in the supernova ejecta.

Gaia17aeq (*Class*: YSO): The comment on the alert page is "long term decline in brightness, near YSO 2MASS J05101100-0328262 (0.11 arcsec), and ASASSN -13db". The calcium emission in Figure 11 supports the YSO classification. It shows how calcium emission lines in the RVS spectra disappear as the YSO gets fainter and much redder, as is evident from the BP spectra becoming weaker compared to the RP spectra. Gaia17aeq is the alert with the highest number of epochs with RVS spectra (seven).

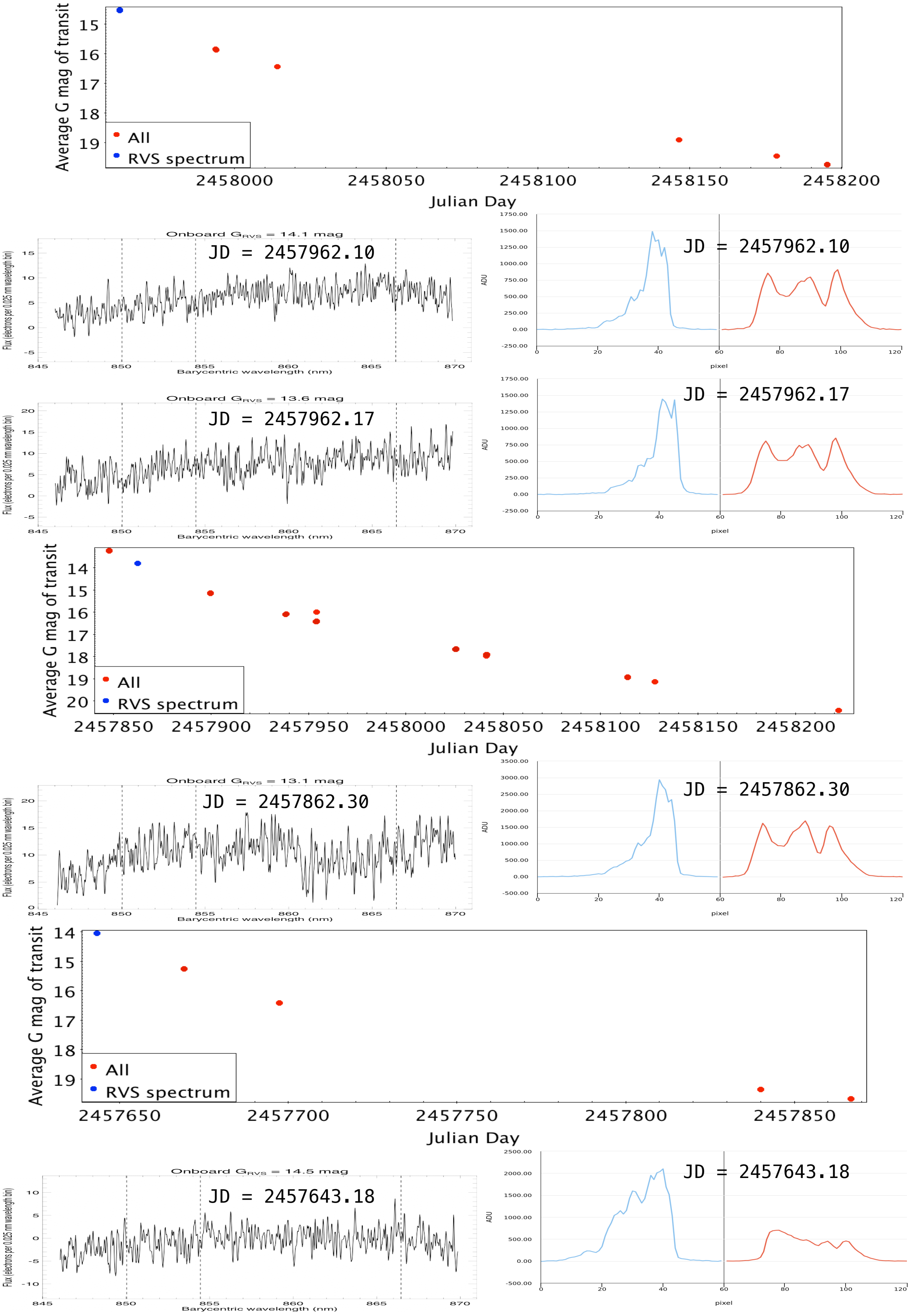

**Figure 8**: Same as Figure 1 but for Gaia17bxg (*Class*: SN Ia, top five panels), Gaia17axt (*Class*: SN Ia, middle three panels), Gaia16bgk (*Class*: SN Ia, bottom three panels).

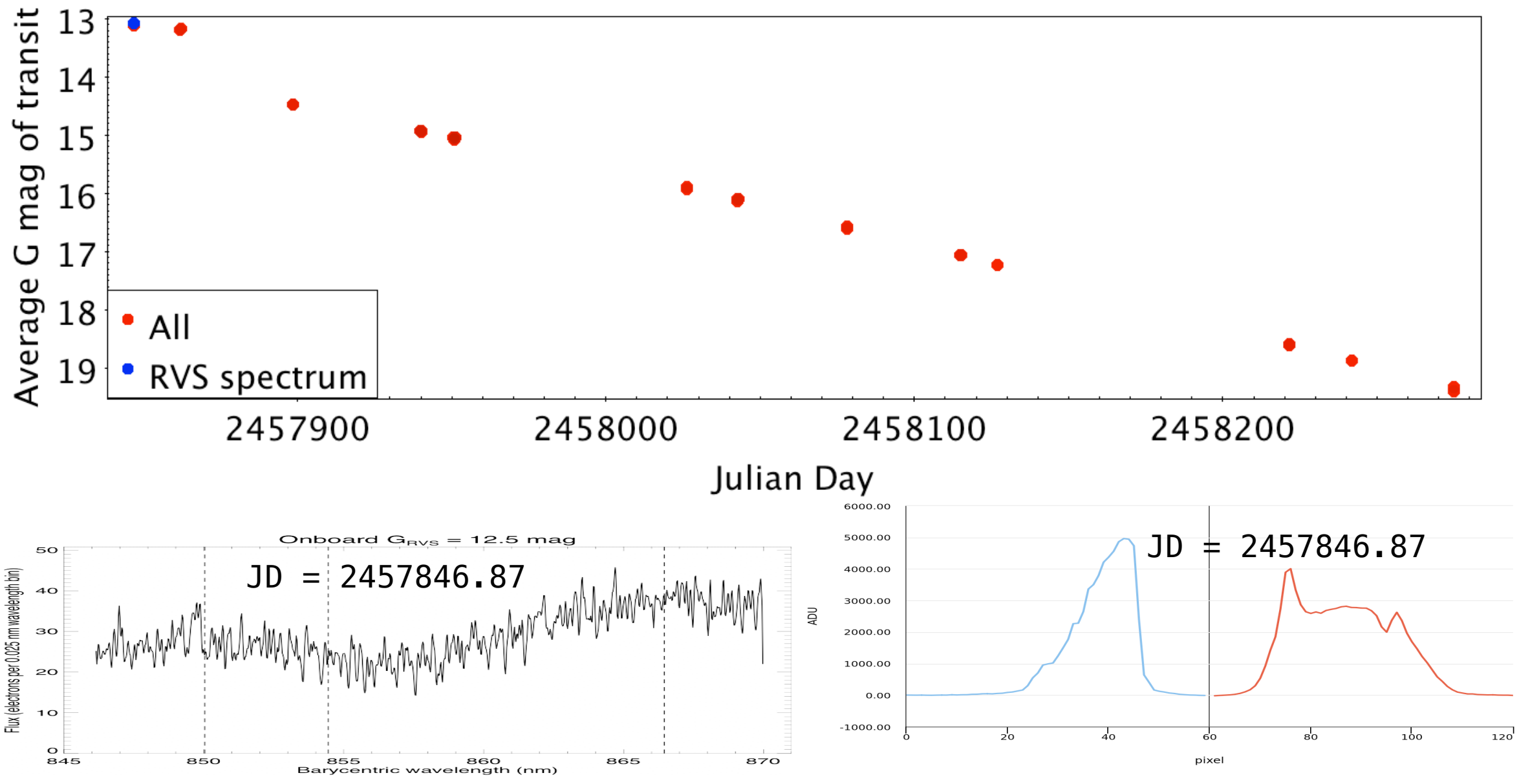

**Figure 9**: Same as Figure 1 but for Gaia17axu (*Class*: SN II).

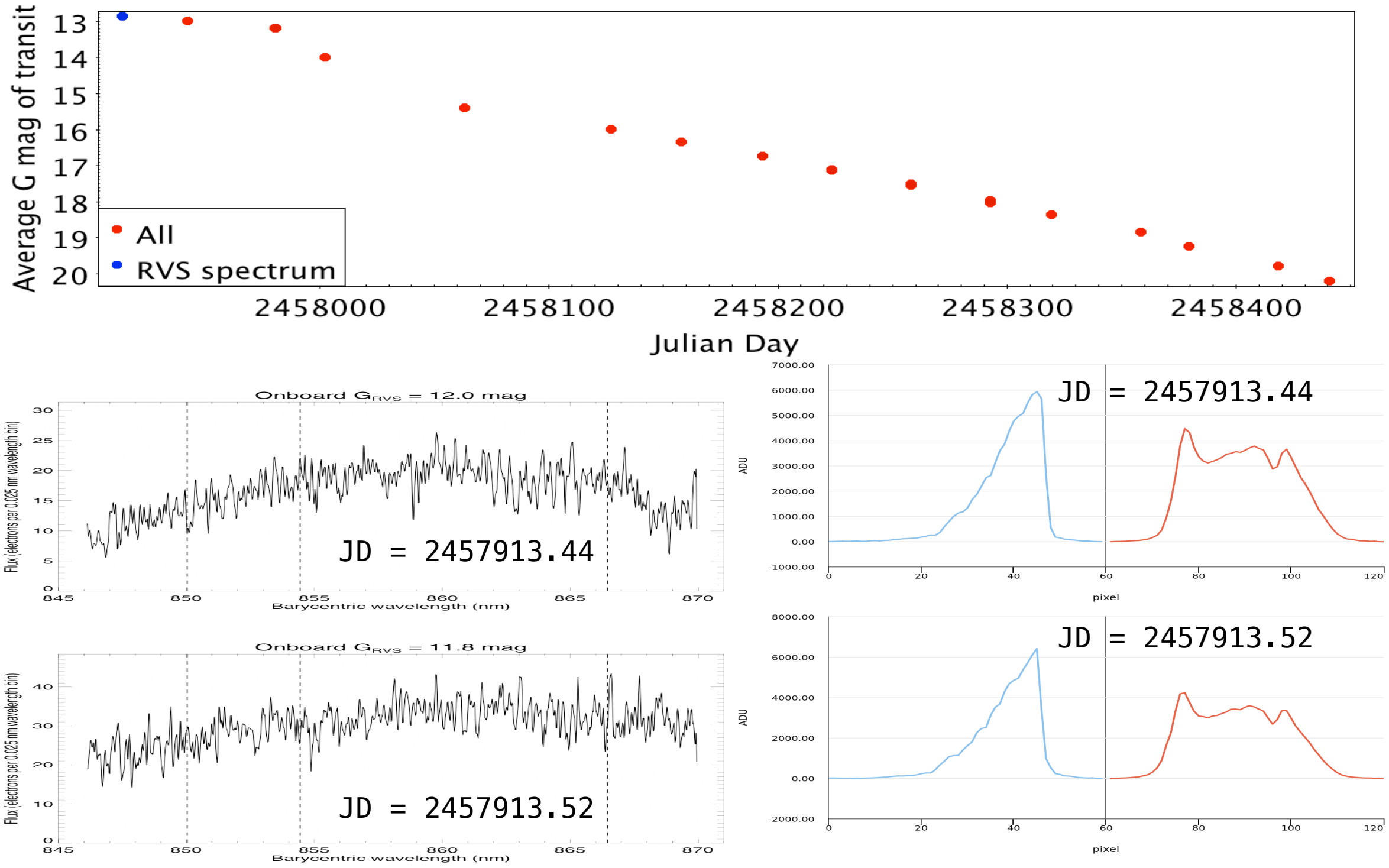

**Figure 10**: Same as Figure 1 but for Gaia17bmy (*Class*: SN IIP).

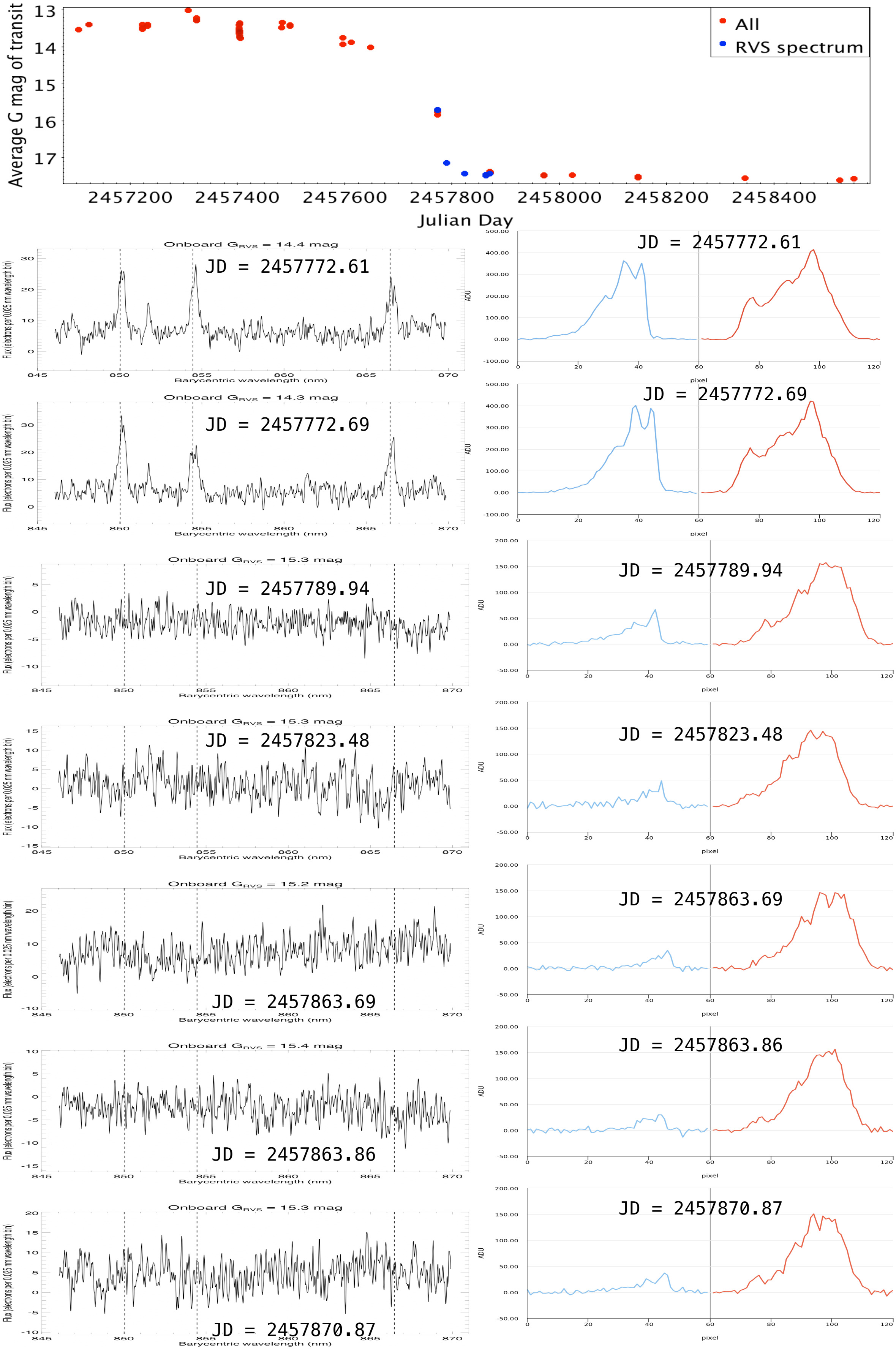

**Figure 11**: Same as Figure 1 but for Gaia17aeq (*Class*: YSO).

## 5. Summary

We present 28 of Gaia's Radial Velocity Spectrometer (RVS) epoch spectra next to their corresponding Gaia Blue Photometer (BP) and Red Photometer (RP) epoch spectra of 13 Gaia Photometric Science Alerts (GPSA) in a single place for the first time. Five of the GPSAs are photometrically classified as unknown. It is hoped this paper will stimulate interest in these sources to help classify them.

The eight photometrically-classified GPSAs are five supernovae (three SN Ia, one SN II, one SN IIP), one catalysmic variable, one binary microlensing event and one young stellar object. These are a preview of the epoch RVS spectra that will be published in Gaia's fourth data release.

## 7. Acknowledgements

This work has made use of data from the European Space Agency (ESA) mission Gaia (https://www.cosmos.esa.int/gaia), processed by the Gaia Data Processing and Analysis Consortium (DPAC, https://www.cosmos.esa.int/web/gaia/dpac/consortium). Funding for the DPAC has been provided by national institutions, in particular the institutions participating in the Gaia Multilateral Agreement. The UK Gaia grant is funded by the UK Space Agency. We acknowledge ESA Gaia, DPAC and the Photometric Science Alerts Team (http://gsaweb.ast.cam.ac.uk/alerts).